\documentclass[11pt]{article}
\usepackage{fullpage} 
\usepackage{algorithmic}
\usepackage{algorithm}
\usepackage{epsfig}

\newcommand{\sr}[2]{\ensuremath{s_{#1,#2}}}
\newcommand{\si}[2]{\ensuremath{s'_{#1,#2}}}
\newcommand{\ri}[2]{\ensuremath{r_{#1,#2}}}
\newcommand{\z}[1]{\ensuremath{P_{#1}}}
\newcommand{\p}[3]{\ensuremath{p^{#1}_{#2,#3}}}

\begin{document}
\title{Inferring Attitude in Online Social Networks Based On Quadratic Correlation}
\author{
\date{} 
Cong Wang\\
Simon Fraser University\\
\texttt{cwa9@sfu.ca}\\
\and
Andrei A. Bulatov\\
Simon Fraser University\\
\texttt{abulatov@sfu.ca}
}

\bibliographystyle{amsplain}
\maketitle

\begin{abstract}
The structure of an online social network in most cases cannot be described 
just by links between its members. We study online social networks, in which 
members may have certain attitude, positive or negative toward each other, 
and so the network consists of a mixture of both positive  and negative 
relationships. Our goal is to predict the sign of a given relationship based on
the evidences provided in the current snapshot of the network. More precisely, 
using machine learning techniques we develop a model that after being trained 
on a particular network predicts the sign of an unknown or hidden link. The 
model uses relationships and influences from peers as evidences for the guess, 
however, the set of peers used is not predefined but rather learned during the 
training process. We use quadratic correlation between peer members to train 
the predictor. The model is tested on popular online  datasets such as Epinions, 
Slashdot, and Wikipedia. In many cases it shows almost perfect prediction accuracy. 
Moreover, our model can also be efficiently updated as the underlaying social 
network evolves.

\end{abstract}

\section{Introduction}
Online social networks provide a convenient and ready to use model of 
relationships between individuals. Relationships representing a wide range 
of different social interactions in online communities are useful for 
understanding and analyzing individual attitude and behaviour as a part of a 
larger society. 

While the bulk of research in the structure on social networks 
tries to analyze a network using the topology of links (relationships) in the network
\cite{Newman}, relationships between members of a network are much 
richer, and this additional information can be used in many areas of social
networks analysis. In this paper we consider signed social networks, which 
consist of a mixture of both positive and negative relationships. This type of
networks has attracted attention of researchers in different 
fields~\cite{Cartwright, Harary, Nowell}. Understanding the interplay of 
relationships of different signs in the online social network setting is crucial for 
the design and function of many social computing applications, where we 
concern about the attitude of their members,
trust or distrust they feel, known similarities and dissimilarities. For example, 
recommending new connections to their users is a common task in many 
online social networks. Yet, without the understanding of the type of relationships, 
an enemy can be introduced to a user as a friend. This framework is also
quite natural in recommender systems \cite{Avesani05:trust,Sarwar00:analysis} 
where we can exploit 
similarities as well as dissimilarities between users and products.

Over the last several years there has been a substantial amount of work 
done studying signed networks, see, 
e.g.~\cite{LeskovecHK10,ChiangNTD11,GuhaKRT04,BurkeK08,Kunegis09}.  
Some of the studies focused on a specific 
online network, such as Epinions~\cite{GuhaKRT04,Massa05:controversial}, 
where users can express trust or distrust to others, 
a technology news site Slashdot~\cite{Kunegis09,Lampe07:follow},  
whose users can declare others `friends' or `foes', and 
voting results for adminship of Wikipedia~\cite{BurkeK08}. Others develop 
a general model that fits several different networks~\cite{ChiangNTD11,LeskovecHK10}.
We build upon these works and attempt to combine the best in the two 
approaches by designing a general model that nevertheless can be tuned up
for specific networks.

\paragraph{Edge sign prediction}
Following Guha et al.~\cite{GuhaKRT04} and  Kleinberg et al.~\cite{LeskovecHK10},
\cite{Nowell} 
we consider a signed network as a directed (or undirected) graph, every edge
of which has a sign, either positive to indicate friendship, support, approval, or
negative to indicate enmity, opposition, disagreement. The edge sign prediction 
problem, in which given a snapshot of the signed network, the goal is to 
predict the sign of a given link using the information provided in the snapshot. 
Thus, the edge sign problem is similar to the much studied link prediction 
problem~\cite{Liben-Nowell03:link,Hasan06:linkprediction}, only we need to 
predict the sign of a link rather than the link itself. 

Several different approaches have been taken to tackle this problem. 
Kunegis et al.~\cite{BrzozowskiHS08} studies the friends and foes on 
Slashdot using network characteristics such as clustering coefficient, centrality 
and PageRank; Guha et al.~\cite{GuhaKRT04}  used propagation algorithms 
based on exponentiating the adjacency matrix to study how trust and distrust 
propagate in Epinion. Later Kleinberg et 
al.~\cite{LeskovecHK10} took a machine learning approach to identify 
features, such as local relationship patterns and degree of nodes, and their
relative weight and build a general model to predict the sign of a 
given link. They would train their predictor on some dataset, to learn the weights 
of these features by logistic regression. Once trained, the model can be used on 
different networks. Clearly, one of the most important measures of an approach
is the accuracy of prediction it provides. Remarkably, in many cases (where comparable)
the network independent approach from \cite{LeskovecHK10} provides 
more accurate predictions than that of previous network specific studies. This
shows certain potential of machine learning techniques. Interestingly, this study is 
also related to the status and balance theories from social psychology 
\cite{Cartwright,Heider}, as they rely on configurations similar to the features
exploited in \cite{LeskovecHK10}.

In this paper we also take the machine learning approach, only instead of focusing 
on a particular network or building a general model across different networks, 
we build a model that is unique to each individual network, yet can be trained 
automatically on different networks. Such an approach intuitively should be capable 
of more accurate predictions than network independent methods, and it remain 
practically feasible. 
 
\paragraph{Trusted peers and influence}
The basic assumption of our model is that users' attitude can be determined by 
the opinions of their peers in the network (compare to the balance and status 
theories \cite{Cartwright,Heider} from social psychology discussed in 
\cite{LeskovecHK10}). Intuitively speaking, peer opinions are guesses from peers 
on the sign of the link from a source node to a target node. Also, we assume that 
peer opinions are only partially known, some of them are hidden. 
We introduce three new components into the model: set of trusted peers, 
influence, and quadratic correlation technique.

When we try to count on peer opinions, not all such opinions are equally 
reliable, and we therefore choose a set of trusted peers whose 
opinions are important in determining the user's action. The set of 
trusted peers is one of the features our algorithm learns during the 
training phase.
Ideally, it would be good to have a set of trusted peers for each link in the 
network. However, considering 
the sparsity and the enormous size  of the network, we cannot always afford to 
determine a set of trusted peers for every possible relationships. Instead, we 
find a set of trusted peers for each individual node. The optimal composition 
of such a set is not quite trivial, because even trusted peers may disagree, and 
sometimes it is beneficial to have trusted peers who disagree. Thus, to make 
reliable estimations on all relationships starting at the individual nodes, its 
set of trusted peers has to form a wide knowledge base on other nodes in 
the network.

While peer opinions provide very important information, this knowledge is sometimes incomplete. 
Relying solely on peer opinions implies that the attitude of a user would always agree with the 
attitude of a peer. However, in reality, there are often exceptions. What also 
matters is how this opinion correlates with the opinion of the user we are 
evaluating. To take this correlation into account we introduce another feature 
into the model, influence. Suppose the goal is to learn the sign of the link between 
user $A$ and user $B$, and $C$ is a peer of $A$. Then if $A$ tends to disagree with 
$C$, then positive attitude of $C$ towards $B$ should be taken as indication that $A
$'s attitude towards $B$ is less likely to be positive. The opinion of $C$ is then 
considered to be the product of his attitude towards $B$ and his influence on $A$.
Usually, influence is not given in the snapshot of the network. For example in the
Wikipedia adminship dataset the explicit information is a collection of results of voting, while the
correlation between the ways members vote is hidden and has to be learned
together with other unknown parameters. We experimented with different ways of 
defining peer opinion, and found that using relationships and influences together to 
approximate peer opinions is more effective than using relationships along.  

To learn the weights of features providing the best accuracy we have 
chosen to use the standard quadratic correlation technique from machine 
learning \cite{guyon06:feature}. This method involves finding the optimum of a quadratic 
polynomial, and while being relatively computationally costly, tends to provide 
very good accuracy. Therefore to solve quadratic programs we resort to 
three approaches. Firstly, we used an available Max-SAT solver METSLib 
\cite{metslib} based on the Tabu search heuristics. Secondly, we also attempted to find 
the exact optimal solution using the brute force approach. Third, we use 
off-the-shelf solver Cplex \cite{cplex}. Clearly, in the latter two approaches 
it is not feasible to solve the quadratic 
program arising from a large network, therefore we also used a number of
heuristics to split such a program, as described later. 
An interesting use of our approach is to apply the quantum annealing 
devise developed by D-Wave \cite{D-Wave} to run our algorithm. This 
devise solves large
instances of the Quadratic Unconstrained  Binary Optimization problem 
(QUBO) with (supposedly) high accuracy and high speed. However,
such experimentation is yet to be done because the device is currently 
unavailable for experimenting. 

\paragraph{Comparison to other work}
Similar to \cite{LeskovecHK10,ChiangNTD11} we also use a machine 
learning approach to build a prediction model based on local features. 
However, unlike their generalized features, such as the degree of 
nodes, and local relationship patterns, we use peer opinions from 
trusted peers which are personalized features. There are two main 
advantages for using personalized features. First of all, our model 
tolerates differences in individual personalities. Unlike existing approach, 
two nodes with the same local features can behave differently by 
selecting different sets of trusted peers. Yet the model of Kleinberg et al.\ 
\cite{LeskovecHK10} treats nodes with the same feature values as the same. Secondly, 
our model accommodates the dynamic nature of online social networks. 
Personalized features allow us to train a predictor separately for each 
individual node. As the network evolves over time, we only need to update 
individual predictors separately instead of rebuilding the whole model. 
Although our model does not generalize across different datasets, a new 
model can be easily trained for different datasets without changing the 
algorithm. 

We build and test our model on three different datasets studied before, 
Epinions, Slashdot and Wikipedia. It is difficult, however, to compare
our results against the results in other works such as 
\cite{GuhaKRT04,BrzozowskiHS08,LeskovecHK10,ChiangNTD11}. 
For example \cite{GuhaKRT04} and \cite{LeskovecHK10} use certain
(different!) normalization techniques to eliminate the bias of the 
datasets above toward positive links. We therefore tried to test 
our model in all regimes used in the previous papers. The results 
shows similar or better prediction accuracy in almost all cases. 
When tested on unchanged (biased) datasets our model shows nearly 
perfect prediction. 
In spite of this, fair comparison is still problematic, because of the lack
of data about other approaches. For example, even 
the experiment results show that our model has 
a better prediction accuracy than the model in ~\cite{ChiangNTD11} 
statistically, we couldn't simply conclude our model is better because 
they used some normalization technique on the dataset, and also, 
it was not specified which edge embeddedness threshold (widely
used in \cite{LeskovecHK10}) was used for the experiment.   

Also, we test the model on a different dataset, MovieLens \cite{movielens}, 
used to recommend users movies to rent. Experiments show that we 
achieve good prediction accuracy on this dataset as well.

\section{Approach}

We now describe our method. We start with the underlying model of a network,
then proceed to the machine learning formulation of the edge sign prediction 
problem, and finally describe the method to solve the resulting quadratic optimization problem.

\subsection{Underlying Model}

We are given a snapshot of the current state of a network.
A snapshot of a network is represented by a directed graph $G=(V,E)$, 
where nodes represent the members of the network and edges the known 
or explicit links (relationships). Some of the links are signed to indicate 
positive or negative relationships. Let $\sr{a}{b}$ denote the sign of the 
relationship from $a$ to $b$ in the network. It may take two different values, 
$\{-1,1\}$, indicating negative and positive relationships respectively. 
Note that nodes of $G$ may represent entities of different kinds. For 
example, a signed relationship can also indicate the like or dislike of a 
product from a user, or the vote from a voter to a candidate. 

To estimate the sign $\sr{a}{b}$ of a relationship from $a$ to $b$, we 
collect peer opinions. By a peer we understand a node in the network
that we use to estimate $\sr{a}{b}$. In different versions of the models
a peer can be any node of the network, or any node linked to $a$. 
Peer opinion is an important unknown parameter 
of the model. It is an estimation on the type of the relationship from a 
peer based on its own knowledge. Let $\p{c}{a}{b} \in \{-1,0,1\} $ denote 
the peer opinion of peer $c$ on the sign $\sr{a}{b}$. When 
$\p{c}{a}{b}=1$ or  $\p{c}{a}{b}=-1$, it indicates that the $c$ believes 
that $\sr{a}{b}=1$ or $\sr{a}{b}=-1$ respectively. When $\p{c}{a}{b}=0$ 
that means $c$ does not have enough knowledge to make a valid estimation.

Another assumption made in our model is that not every peer can make 
a reliable estimation. Therefore we divide all peers of a node into two 
categories, and count the opinions only of the peers from the first category,
trusted peers. The problem of
how to select a set of trusted peers and use their opinions for the estimation
will be addressed later. 
Let $\z{a}$ denote the set of trusted peers of $a$. We estimate the sign 
$\sr{a}{b}$ of a relationship from $a$ to $b$ by collecting the opinions of 
peers $c \in \z{a}$. If the sum of the opinions is nonnegative, then we say 
$\sr{a}{b}$ should be $1$, otherwise, it should be $-1$. This can be 
expressed by a simple equation as,
\begin{equation}
\sr{a}{b}=sign(\sum_{c \in \z{a}}\p{c}{a}{b})
\end{equation}

Notice that the set of trusted peers $\z{a}$ for each node $a$ is also 
an unknown parameter. Determining the set of nodes in the set of trusted 
peers is a nontrivial task. Observe, for example, that the prediction 
accuracy does not necessarily increase as we add nodes, even nodes of higher 
trust into the set. Since the estimation are made by collecting opinions 
from all peers in the set, a correct estimation from one peer can be 
canceled by the wrong estimation of another. Also, it is beneficial to
select a set of peers with more diversity without compromising accuracy. 
As mentioned earlier, the set of trusted peers  of a node $a$ is crucial 
for the estimation of all relationships starting from $a$. Hence, having a 
set of peers that make good individual estimation on relationships to 
different sets of target nodes rather than nodes that make good individual 
estimation for the same set of target nodes will likely improve the accuracy 
of prediction.     

\subsection{Machine learning approach}

Our approach to selecting an optimal set of trusted peers is to consider 
the quadratic correlations between each pair of peers. The overall 
performance of a set of peers is determined by the sum of the individual 
performance of each of them together with the sum of  their performance 
in pairs. The 
individual performance measures the accuracy of individual estimations, 
while the pairwise performance measures the degree of difference 
between the estimations of the pair of peers. We want to maximize the 
accuracy of each individual and the diversity of each pair at the same time. 


\paragraph{The loss function}
Our goal is to use the information in $G$ to build a predictor $S(x,y)$ that 
predicts the sign $\sr{x}{y}$ of an unknown relationship from $x$ to $y$ with 
high accuracy. At the same time, we would also determine the unknown 
parameters used in the model. Our goal can be expressed by the objective 
function below,
\begin{equation} \label{eq:00}
min_S\{\sum_{x,y}(S(x,y)-\sr{x}{y})^2\}.
\end{equation}
Least square loss function is the standard loss function used 
in measuring the prediction error. Another important reason for us to pick 
the square loss function is that it helps to  capture quadratic correlations 
between all pairs of nodes in $V$. The quadratic correlation becomes more 
clear when the term gets expanded later in the next section. 

Function $S(x,y)$ is defined as the sign of the sum of peer 
opinions as follows. Let  
\begin{equation}\label{eq:0}
F_x(y) = \sum_{z \in \z{x}}\p{z}{x}{y} 
\end{equation}
denote the sum of individual peer opinions. Then we set
\begin{equation} 
S(x,y) = \left\{ 
  \begin{array}{l l}
    1 & \quad \mbox{if $F_x(y)\geq 0$ }\\
    -1 & \quad \mbox{if $F_x(y) < 0$}\\
  \end{array} \right.
\end{equation}

Since $\z{x}$ is unknown, we introduce a new variable 
$w_{z,x} \in\{0,1\}$ which indicates if a node $z \in V$ should be 
included into set $\z{x}$. Hence, we rewrite Equation~(\ref{eq:0}) 
using the characteristic function $w_{z,x}$ as, 
\begin{equation}\label{eq:01}
F_x(y)=\sum_{z \in V}w_{z,x}\p{z}{x}{y} 
\end{equation}

\paragraph{Quadratic optimization problem}
We are now ready to set the machine learning problem. A
training dataset (a subset of $G$) is given. Every entry of the training dataset 
is a known edge along with its sign. Let a training dataset be 
$D=\{(x_i, y_i,\sr{x_i}{y_i})| i=1, ... , M\}$. The goal is to minimize the objective 
function, finding the optimal weight vector $w=\{w_x | x \in V \}$, where 
$w_x=\{w_{z,x}|z \in V\}$. We use machine learning methods 
\cite{guyon06:feature} to train the 
predictor $S(x,y)$ and learn an optimal weight vector such that the objective 
function~(\ref{eq:00}) is minimized. Substituting $S(x,y)$ and $F_x(y)$ into 
the objective function~(\ref{eq:00}) we obtain  
quadratic unconstrained binary optimization (QUBO) problem 
described by equation~(\ref{eq:2}).
\begin{equation}\label{eq:2}
w^{opt} = arg min_{w}\left( \sum_{(x,y,\sr{x}{y})\in D}(\frac{1}{N}
\sum_{z\in V}(w_{z,x}\p{z}{x}{y}-\sr{x}{y})^2)\right)
\end{equation}

We want to minimize the amount of error made by $S(x,y)$, yet at the 
same time, we could also want to avoid overfitting. So we introduce the 
second term which is the regularization function based on the $L0$-norm. 
It ensures that the size of $\z{x}$ is not too large. The parameter $\lambda$ 
controls the trade off between the accuracy of the prediction and the 
complexity of the set $\z{x}$. Thus, the final form of the objective 
function is as follows
\begin{equation}\label{eq:2a}
\hspace*{-5mm}
w^{opt} = arg min_{w}\left( \sum_{(x,y,\sr{x}{y})\in D}\left(\frac{1}{N}
\sum_{z\in V}(w_{z,x}\p{z}{x}{y}-\sr{x}{y})^2+\lambda|w|\right)\right)
\end{equation}

Note that there will 
be more details on peer opinion terms $\p zxy$. 

\subsection{Peer opinion variants}
As mentioned earlier, we are going to test our model using different 
peer opinion formulations. First, let $\si xy$ be extension of $\sr xy$ 
to edges with unknown sign and also to pairs of nodes that are not
edges defined by
$$
\si xy=\left\{\begin{array}{ll}
\sr xy & \mbox{if $\sr xy$ exists},\\
0, & \mbox{otherwise}.
\end{array}\right.
$$

\paragraph{Simple-adjacent}
The simplest option, later referred to as 
\emph{Simple-adjacent}, is, based on the given information, to formulate 
peer opinions using existing relationships from peers to the target node. 
In other words, in this case we set $\p zxy$ to be $\si{z}{y}$. 

\smallskip

However, we also understand that the relationship from a peer to the 
target node does not always agree with the relationship from the source 
node to the target node. Yet, peers whose attitudes always disagree 
with the sources node are as important as these whose attitudes agree 
with the source node. In order to take the advantage of these disagreements, 
we introduced a second parameter, influence, which can be either positive, 
negative, or neutral.  

\paragraph{Standard-pq}
The second and third options differ in who is considered as a peer.
In the \emph{Standard-pq} option the influences $\ri{x}{y} \in \{-1, 0, 1\}$ 
associated with each pair of vertices $x,y$ is an unknown parameter in 
the model. In a sense every pair of nodes is assumed
to be linked, thus turning $G$ into a complete graph. A positive influence, 
$\ri{x}{y}=1$, indicates that the attitude 
of $x$ affects $y$ positively, while a negative influence, $\ri{x}{y}=-1$ 
indicates that the attitude of $x$ affects $y$ negatively. 
Then the we obtain the standard formulation, 
\begin{equation}\label{eq:1}
\p{z}{x}{y}=\si{z}{y}\ri{z}{x}
\end{equation}
Since the standard formulation gives us the best result in experiments, 
we use it throughout our discussion. Using the standard formulation, 
we rewrite Equation~(\ref{eq:01}) as
\begin{equation}\label{eq:02}
F_x(y)=\sum_{z \in V}w_{z,x}\si{z}{y}\ri{z}{x}. 
\end{equation}

\paragraph{Standard-adjacent}
Finally, in the \emph{Standard-adjacent} option peers of $x$ are restricted 
to the neighbours of $x$. 
$$
F_x(y)=\sum_{z \in N(x)}w_{z,x}\si{z}{y}\ri{z}{x}. 
$$
The rest is defined in the same way as for the 
\emph{Standard-pq} option.

\subsection{Simplifying the model}
In our model, we are given a directed complete graph $G=(V,E)$.  
In Equation~(\ref{eq:02}), both $w_{z,x}$ and $\ri{z}{x}$ are unknown 
parameters. Since $\ri{z}{x}\in\{-1,0,1\}$, we can reduce the number of 
unknown parameters by considering all possible values of $\ri{z}{x}$, and 
rewriting $F_x(y)$ as,
\begin{equation}\label{eq:03}
F_x(y)=\sum_{z \in V}w_{z,x}^{+}\si{z}{y}-w_{z,x}^{-}\si{z}{y}
\end{equation}
where $w_{z,x} = w_{z,x}^{+}+ w_{z,x}^{-}$ for 
$w_{z,x}^{+}, w_{z,x}^{-} \in \{0,1\}$. If $w_{z,x}^{+}=1$, then 
$z \in \z{x}$ and $\ri{z}{x}=1$. Similarly, $w_{z,x}^{-}=1$ indicates that
$z \in \z{x}$ and  $\ri{z}{x}=-1$. When both $w_{z,x}^{-}=0$ and 
$w_{z,x}^{+}=0$, then $z \not\in \z{x}$. Although $\ri{z}{x}$ can take 
three possible values, there are only two terms in Equation~(\ref{eq:02}) 
since when $\ri{z}{x}=0$, the term is also zero regardless of the value 
of $\si{z}{y}$. 

Now to minimize the objective function, we need to determine the optimal 
weight vector $w=\{w_x | x \in V \}$ where 
$w_x=\{w_{z,x}^+,w_{z,x}^-|z \in V\}$ such that
\begin{equation}\label{eq:20}
w^{opt} = arg min_{w}\left( \sum_{(x,y,\sr{x}{y})\in D}
\left(\frac{1}{N}\sum_{z\in V}(w_{z,x}^+-w_{z,x}^-)\si{z}{y}-
\sr{x}{y}\right)^2+\lambda|w|\right)
\end{equation}
  
To find the optimal solution of $w$, we need to solve a QUBO of $2n^2$ 
variables which is very difficult since there are usually millions of nodes in 
a social network. Luckily, from the definition, we know $w_x$ and $w_y$ 
are independent for different nodes $x$ and $y$. Instead of solving for 
$w^{opt}$ directly, we can solve $w_x^{opt}$ for each $x \in V$ separately, 
and then combine their values to get $w^{opt}=\{w_x^{opt}| x \in V\}$.  
\begin{eqnarray}\label{eq:21}
w_x^{opt} &=& arg min_{w_x}\left( \sum_{(x,y,\sr{x}{y})\in D}
\left(\frac{1}{N}\sum_{z\in V}
(w_{z,x}^+-w_{z,x}^-)\si{z}{y}-\sr{x}{y}\right)^2+\lambda|w_x|
\right)\\ \nonumber
&=&arg min_{w_x} \left(\frac{1}{N^2}\sum_{v \in V}\sum_{u \in V}
(w_{v,x}^+-w_{v,x}^-)(w_{u,x}^+-w_{u,x}^-)\left(
\sum_{(x,y,\sr{x}{y})\in D}\si{v}{y}\si{u}{y}\right)\right.\\ 
& &\ \ +\sum_{z \in V}(w_{z,x}^+-w_{z,x}^-)\left(\lambda -\frac{2}{N}
\sum_{(x,y,\sr{x}{y})\in D}\si{z}{y}\si{x}{y}\right)\Bigg) \nonumber
\end{eqnarray}

Now, instead of solving a QUBO of size $2n^2$, we could solve $n$ QUBOs 
of size $2n$ separately. It can be solved approximately by a Max-SAT
solver

If we use a different approach (similar to \cite{dwave}), the problem 
should be further simplified, as 
it is still challenging to solve each of these size $2n$ QUBOs exactly. 

\paragraph{Breaking down the problem}
In order to find a good approximation of the optimal solution to the QUBO 
defined by Equation~(\ref{eq:21}), we could break it down to much smaller 
QUBOs. Given a subset $U \subset V$, let 
$w_{x,U} = \{w_{z,x}^+,w_{z,x}^-|z \in U\}$, and define a restricted 
optimization problem as follows
\begin{equation}\label{eq:22}
w_{x,U}^{opt} = arg min_{w_{x,U}}\left( \sum_{(x,y,s_{x,y})\in D}
\left(\frac{1}{N}\sum_{z\in U}(w_{z,x}^+-w_{z,x}^-)s_{z,y}-s_{x,y}\right)^2
+\lambda|w_{x,u}|\right)\nonumber
\end{equation}

By the definition when $U=V$, Equation~(\ref{eq:22}) is exactly 
the same as ~(\ref{eq:21}), However, if we allow $U$ to be a subset of much 
smaller size, then the QUBO we need to solve is a much smaller size as well. 
By decreasing the size of $U$, we are trading the solution accuracy with the 
computational efficiency. Our intuition is to arrange nodes in $V$ according to 
some  order, and then, we break $V$ into several smaller subsets. 
By solving Equation~(\ref{eq:22}) on each of the subsets $U$, and combining 
their solutions, we can get a good approximation to the optimal solution of 
Equation~(\ref{eq:21}). In the next section, we are going to explain the method 
used to approximate the optimal value of $w_x$ in detail. 
        
\subsection{Method}
The optimization problem defined by Equation~(\ref{eq:21}) is a quadratic 
unconstrained binary optimization problem which is NP-hard in general. To solve 
the optimization problem, we use two approaches. First, we solve problems
of the form (\ref{eq:21}) separately using an open source  Max-SAT solver 
METSLib based on the Tabu search heuristics. Second, we apply a similar 
method as described in~\cite{dwave} to reduce the size of the problem dramatically. 

Under the second approach we break the original problem into even 
smaller subproblems and we obtain an approximate solution by combining the 
solutions of each subproblems. Algorithm~\ref{alg:1} describes the method we 
used to solve each subproblems. Algorithm~\ref{alg:2} uses Algorithm~\ref{alg:1} 
as a subroutine and explains how the problem is broken down into subproblems 
and also how to combine the solutions of subproblems to obtain an approximate 
solution. During the training process, we need to use graph $G$ and dataset $D$. 
Moreover, $D$ is randomly split into two parts, training dataset $TD$ and 
validation dataset $VD$. We use $TD$ to train the predictor $S(x,y)$ and we use 
$VD$ to validate the predictors obtained at each step to select the optimal one. 

According to Algorithm~\ref{alg:2}, we split $V$ into smaller subsets in two steps. 
First of all, for each $a\in V$ and for each $v \in V$, we compute the individual prediction error of $v$ for $a$ on
the dataset $TD$ as follows: For each data point $(a,u,\sr{a}{u})\in TD$, we count $e_{v_-}$, the number of instances $\p{v}{a}{u} \ne \sr{a}{u}$ when $\ri{v}{a}=-1$, and $e_{v_+}$, the number of instances $\p{v}{a}{u} \ne \sr{a}{u}$ when $\ri{v}{a}=1$ separately. Note that since $\p{v}{a}{u}=\ri va \si vu$, we can compute this number; and that if $u$ is not a neighbour of $v$ then it contributes to both $e_{v_-}$ and $e_{v_+}$. Then, we replace $v$ by $v_+$ and $v_-$ with individual prediction error, $e_{v_+}$ and $e_{v_-}$ respectively.   
Second, using their individual prediction errors, 
we can sort the nodes in $V$ in increasing order of the error. The subset $U$ is 
iteratively selected by picking the first $d$ nodes in the list that are not yet considered. 
The value of $d$ is an important parameter of the algorithm and is selected 
manually at the beginning of the algorithm. If $d=n$, 
then we are solving the problem defined by Equation~(\ref{eq:21}).  The sorting and 
selecting processes not only reduce the amount of computation, but also allow us to 
consider the relevant nodes first.        

Once the subset $U$ is selected, we use Algorithm~(\ref{alg:1}) to solve the 
subproblem defined by Equation~\ref{eq:22}. The algorithm determines the optimal 
value of $w_{x,U}$ and the set $\z{x} \in U$ which minimize the amount of prediction 
errors made by $S(x,y)$. It repeatedly solves the QUBO for different 
$\lambda \in [\lambda_{min}, \lambda_{max}]$ (in our experiments we use $\lambda_{min}=0.1$ and $\lambda_{max}=0.35$). $\lambda_{min}$ and 
$\lambda_{max}$ bound the possible range for $\lambda$, and the best value of 
$\lambda$ is selected using cross-validation on dataset $VD$. The set $\z{x}$ 
together with the weight $w_{x,U}$ which produces the lowest prediction errors on 
$VD$ is selected as the optimal solution. Notice, when the size of $U$ is small, say 
$d=10$, we can solve it exactly using brute-force method. When its size is larger, 
we use some heuristic methods such as the quadratic optimization solver Cplex 
\cite{cplex} to approximate the solution.     

The optimal solution determined by Algorithm~\ref{alg:1} on subset $U$ is 
used to extend the optimal solution for $w_x$ by Algorithm~\ref{alg:2}. 
To extend the solution, we use a greedy approach similar to AdaBoost~\cite{ada}. 
We would extend the partial solution, as long as extending it by the optimal 
solution on $U$ lowers the prediction error on the validation set $VD$.  

\begin{algorithm}
\caption{Set the parameter for a subset}
\label{alg:1}
\begin{algorithmic}
\REQUIRE training dataset: $TD$, validation dataset: $VD$, a subset of nodes:~$U$
\ENSURE values of $w$ and $Z \subset U$
\STATE $Z = \emptyset$
\STATE $e_{min}$ = $|TD|$
\FOR{$\lambda = \lambda_{min}$ to $\lambda_{max}$} 
\STATE $Z_{current}=\emptyset$
\STATE solve the optimization $w^{opt} = arg min_w(\sum_{i=1}^{M}(\frac{1}{N}\sum_{z\in U}(w_z^+-w_z^-)s_{z,y}-s_{x,y})^2+\lambda|w|)$
\IF{ $w_z$ == 1}
\STATE  $Z_{current}$ =  $Z_{current} \cup z$
\ENDIF
\STATE Measure the validation error $e_{val}$ on $vd$ using $Z_{current}$
\IF{ $e_{val} < e_{min}$}
\STATE $Z_{final}$  =$Z$
\STATE $e_{min}$ = $e_{val}$
\ENDIF
\ENDFOR
\end{algorithmic}
\end{algorithm}

\begin{algorithm}
\caption{solve the optimization problem}
\label{alg:2}
\begin{algorithmic}
\REQUIRE training dataset: $TD$, validation dataset: $VD$, The size of the subset: $d$ 
\ENSURE values of $w_z$ for $z \in V$, and the set of trusted peers $Z$
\STATE $e_{old}$ = $|TD|$
\STATE $e_{new}$ =$|TD|$-1
\STATE $Z=\emptyset$
\STATE $Z_{current}=\emptyset$
\STATE sort nodes of $V$ by their individual prediction errors in increasing order
\STATE $U$ = the first $d$ nodes in $V$ 
\WHILE{$e_{old}>e_{new}$}
\STATE $Z$ =  $Z_{current} \cup Z$
\STATE $w_z$, $Z_{current}$ = Algorithm~\ref{alg:1}($td$, $vd$, $U$)
\STATE $e_{old}$=$e_{new}$
\STATE Measure the validation error $e_{new}$ on $vd$ using $Z$
\STATE update $U$ with the next $d$ nodes in $V$
\ENDWHILE
\end{algorithmic}
\end{algorithm}

\section{Datasets}

We use three datasets borrowed from \cite{LeskovecHK10} and a movie
rental dataset Movielens that we consider separately. In order to make 
comparison possible the datasets are unchanged rather than updated to their 
current status. The dataset statistics is therefore also from \cite{LeskovecHK10}
(see Table~\ref{tab:stat}).

\paragraph{Epinions}
This is a web site dedicated to reviews on a variety of topics including product 
reviews, celebrities, etc. The feature of Epinion interesting to us is that users can 
express trust or distrust to each other, making it a signed social network. 
The dataset contains 119,217 nodes, 841,200 edges, 85\% of which are 
positive and 15\% are negative.

\paragraph{Slashdot}
Slashdot is another web site for technology news where users are allowed to
leave comments. It also has an additional Zoo feature that allows users tag
each other as `friends' and `foes'. This dataset contains 82,144 nodes, 
549,202 edges of which 77.4\% are positive and 22.6\% are negative.

\paragraph{Wikipedia}
This dataset contains Wikipedia users and the results of voting among them 
for adminship. Every link represents a vote of one user for or against another. 
A link is positive if the user voted for another and negative otherwise. 
It contains 7,118 nodes representing users who casted a vote or been voted for, 
103,747 edges, of which 78.7\% are positive and 21.2\% are negative.

\begin{table}
\caption{Basic statistics on the datasets}\label{tab:stat}
\label{tb1}
\begin{tabular}{ |l|c|c|c| }
\hline
Dataset &Epinions&Slashdot&Wikipedia\\ \hline
Nodes & 119217 & 82144 & 7118 \\ \hline
Edges &841200&549202&103747\\ \hline
+1 edges & 85.0\%& 77.4\% & 78.7\%\\ \hline
-1 edges &15.0\% &22.6\% &21.2\%\\
\hline
\end{tabular}
\end{table}

\section{Experiment}

\subsection{Parameters of datasets}

In our experiment, we split each dataset into two parts. We randomly pick one 
tenth of the dataset for testing. The remaining dataset is used for training. 
The training dataset is split into two equal parts, half for training and half for 
validating during the training process. The datasets used in the experiment are 
sparse and unbalanced. As shown in Table~\ref{tb1}, in these datasets 
approximately 80\% of the edges are positive edges. 

When the dataset is sparse, it is hard to build good classifiers due to the lack of 
training and testing data. In order to get a better understanding of the performance 
of the model, edge embeddedness of an edge $uv$ is introduced 
in~\cite{LeskovecHK10,ChiangNTD11} as the number of common neighbours 
(in the undirected sense) of $u$ and $v$.  Instead of testing the model over the 
entire dataset, they only consider the 
performance restricted to subsets of edges of different levels of minimum 
embeddedness. For example, Kleinberg et al.~\cite{LeskovecHK10} restrict the 
analyses to edges with minimum embeddedness 25. 

Similarly, we also introduce two parameters that restricts the analysis of our 
model. First of all, to measure the knowledge of peers, we introduced a parameter 
$q$. We make a prediction on the relationship from $x$ to $y$ if $y$ is connected 
to at least $q$ peers of $x$. This restriction is similar to the edge embeddedness 
used in~\cite{LeskovecHK10,ChiangNTD11}. If we only consider neighbours of $x$ 
as peers, then $q$ is the same as edge embeddedness. If every node is considered 
a peer of $x$ then $y$ is admitted if its degree is at least $q$. Secondly, we also consider 
another restriction on peers. We will only process nodes that have at least $p$ 
common neighbours with $x$ as peers. When $p=0$, we consider every node in 
the network as a peer of $x$. 

In Table~\ref{tb0}, we show the dependence of the prediction accuracy 
of our model using different $p$ and $q$. The data in Table~\ref{tb0} is obtained 
using option \emph{Standard-pq} with $d=10$ solved by Cplex. As $p$ and $q$ 
grows, the performance of the model clearly improves. However, increasing the
values of $p$ and $q$ severly restricts the set of nodes that can be processed.
We choose somewhat optimal values of these parameters, $q=20$ and $p=15$. In the rest of 
our experiments these values are used. It also worth to notice, $q=20$ and 
$p=15$ is less restrictive than edge embeddedness 25. As shown in Table~\ref{tb00}, 
more edges pass the $q=20$ and $p=15$ threshold than the edge embeddedness 25 
threshold.

\begin{table} 
\caption{Number of edges passing the threshold}
\label{tb00}
\begin{tabular}{ |l|c|c|c| }
\hline
Dataset &Epinions&Slashdot&Wikipedia\\ \hline
(p,q)=(15,20) &247725 &25436 &51372 \\
embeddedness 25 &205796 &21780 &28287 \\
\hline
\end{tabular}
\end{table}

\begin{table} 
\caption{Prediction Accuracy for Different Values for $p,q$}
\label{tb0}
\begin{tabular}{ |l|c|c|c| }
\hline
Dataset &Epinions&Slashdot&Wikipedia\\ \hline
(p,q)=(10,0) & 91.7\% & 84.2\% & 85.0\% \\
(p,q)=(10,10) & 92.8\% & 91.6\%  & 86.5\% \\
(p,q)=(10,20) & 93.7\% & 93.9\% & 86.6\% \\
(p,q)=(10,30) & 95.6\% & 95.1\% & 87.6\% \\
(p,q)=(15,0) & 93.7\% & 87.7\% & 85.2\% \\
(p,q)=(15,10) & 95.8\% & 96.1\% & 86.3\% \\
(p,q)=(15,20) & 96.2\% & 97.9\% & 86.9\% \\
(p,q)=(15,30) & 96.3\% & 96.0\% & 88.5\% \\
(p,q)=(20,0) & 93.5\% & 87.4\% & 85.0\% \\
(p,q)=(20,10) & 96.2\% & 98.1\% & 86.8\% \\
(p,q)=(20,20) & 96.3\% & 98.6\% & 86.9\% \\
(p,q)=(20,30) & 96.5\% & 99.2\% & 89.0\% \\
\hline
\end{tabular}
\end{table}

\subsection{Results}
As explained before, our model depends on several parameters: internal 
parameters such as the peer opinion variant and the method of solving the QUBO,
and external parameter such as balancing the dataset. We first make the 
comparison for different settings of the internal parameters. In this case we 
use the original, unbalanced datasets.

\begin{figure*}[ht]
\centering
\epsfig{totalheight=7cm, file=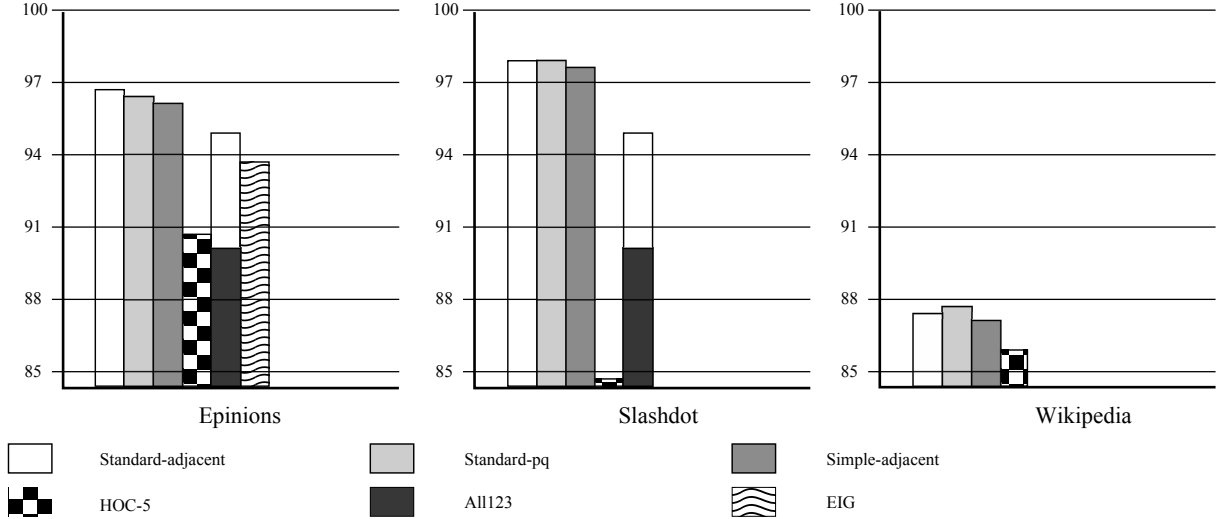}
\caption{Prediction Accuracy of Different Peer Opinion Formulations,
and Previous Results}
\label{fig:formulation}
\end{figure*}

Peer opinion is an important parameter in our model, it can be formulated in 
several different ways. In our experiment, we considered two different formulations, 
the standard formulation which uses both relationships and influences, and the 
simple formulation which uses relationships alone without influences. Moreover, 
we also consider different set of peers, the set of adjacent nodes and the set 
of nodes with at least $p=15$ common neighbours. In Table~\ref{tb3}
(see also Fig.~\ref{fig:formulation}), we compare 
the performance of our model using different peer opinion formulations and peers.  
As shown in the table, standard formulations (\emph{Standard-adjacent}, 
\emph{Standard-pq}) have better 
prediction accuracy then the simple formulation (\emph{Simple-adjacent}), so it is useful 
to introduce influences into the peer opinion formulation. For Slashdot and Wikipedia, 
restricting peers to neighbours (\emph{Standard-adjacent}) is not as effective as using 
the set of nodes with at least $p=15$ common neighbours as peers (\emph{Standard-pq}),
although the difference is neglegible. Surprisingly, 
for Epinions, it is slightly better to only consider neighbours as peers. We compare the 
results with those of \cite{ChiangNTD11} (HOC-5) and \cite{LeskovecHK10} (All123). Unfortunately, 
Kleinberg et al.\  \cite{LeskovecHK10} provide only a (somewhat wide) 
range of the results their model produces on such datasets. However,
even such partial results allow us to conclude that collecting opinions from 
trusted peers is an effective method to infer people's attitude. 

\begin{table} 
\caption{Prediction Accuracy of Different Formulations}
\label{tb3}
\begin{tabular}{ |l|c|c|c| }
\hline
Dataset &Epinions&Slashdot&Wikipedia\\ \hline
Standard-adjacent & 96.59\% & 97.93\% &87.29\%  \\
Standard-pq & 96.36\% &98.00\% &87.69\% \\
Simple-adjacent & 95.93\% &97.68\%&87.03\% \\
HOC-5~(\cite{ChiangNTD11}) & 90.80\% & 84.69\% & 86.05\% \\
All123~(\cite{LeskovecHK10}) & 90-95\% & 90-95\% & N/A \\
EIG (\cite{GuhaKRT04}) & 93.60\% &  N/A & N/A \\
\hline
\end{tabular}
\end{table}

\begin{figure}[ht]
\centerline{\epsfig{totalheight=7cm, file=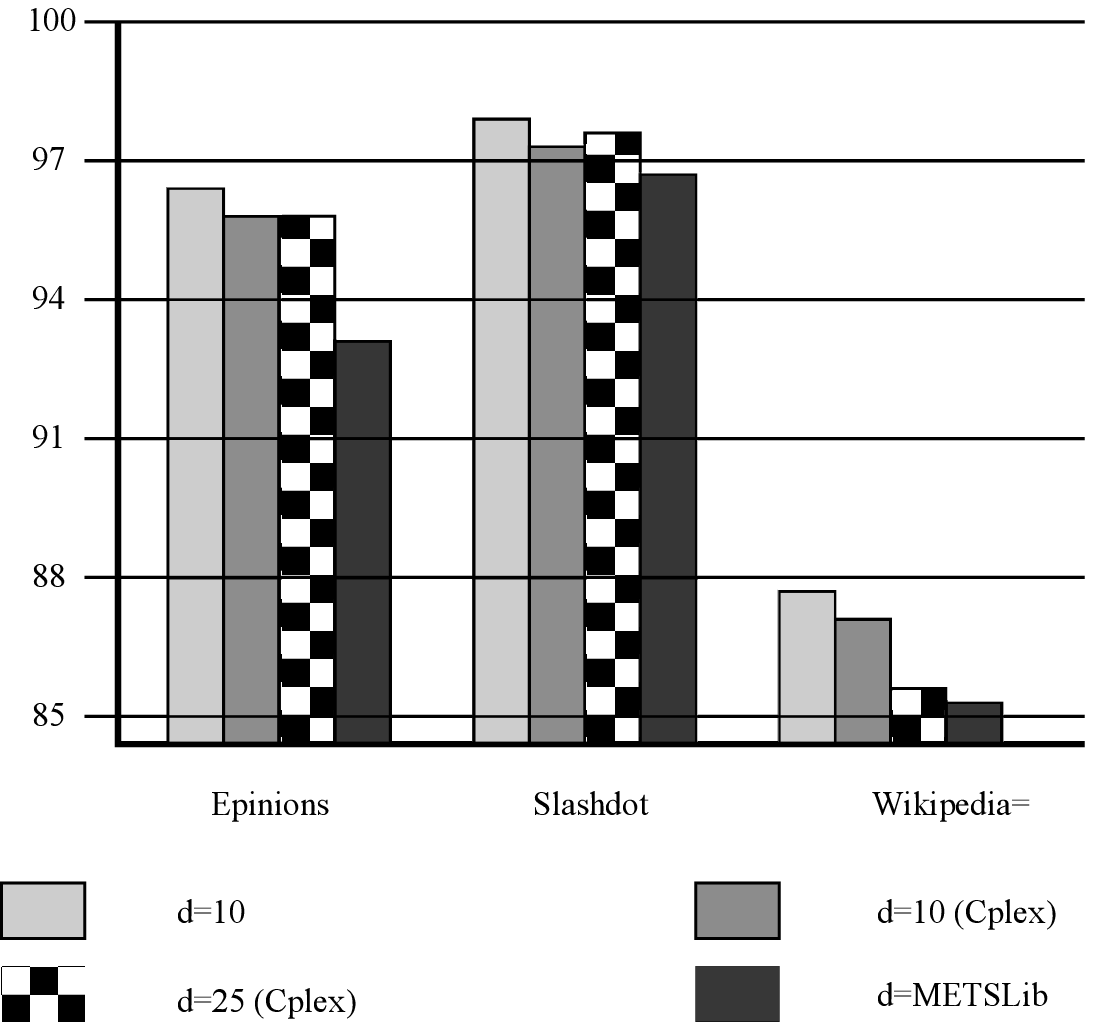}}
\caption{Prediction Accuracy for Different QUBO Solvers}
\label{fig:d}
\end{figure}

As mentioned earlier, solving the quadratic optimization problem is an important 
and the most difficult part of our method. By decreasing the subset size $d$, we 
trade off the accuracy over efficiency. In the experiment, we assign different values 
to $d$ and measure the prediction accuracy. For $d=10$, we can solve the 
optimization problem exactly by brute-force. For $d>10$, we need to use 
Cplex-solver~\cite{cplex} to solve the problem. In Table~\ref{tb4}
(see also Fig.~\ref{fig:d}), we compare 
the performance of our model with different values of $d$. We expect the prediction 
accuracy to increase as the value of $d$ increases. However, experimental result 
shows that it is not the case. As shown in the table, when $d=10$, our model 
gives a better prediction accuracy using exact solver instead of Cplex-solver. 
Since Cplex-solver only gives an approximate solution which effects the overall 
quality of the classifier, we do not see much improvements as $d$ increase.  
But still, we think the prediction accuracy should increase if we can a better 
solution for the problem for larger value of $d$. Indeed, when $d$ equal to the size of 
the original problem, we are solving the original problem directly instead of 
subdividing it into smaller problems. What seems to be the case is that the
accuracy of the algorithm is very sensitive to the quality of approximation 
of the quadratic optimization problem. 

We also include the results obtained by
using the open source Max-SAT solver METSLib \cite{metslib}. Although 
its approximation is clearly inferior to that of Cplex, it solves the entire
problem whiout splitting it into small subproblems. So, the overall 
performance is similar to that of Cplex, except fo Epinion, for which the
time limit set for Tabu search (1 sec) is apparently insufficient. Note as well
that overall running time (although we did not make any precise measurements) 
using METSLib is considerably greater than that using Cplex.
\begin{table} 
\caption{Prediction Accuracy of Different Values of $d$}
\label{tb4}
\begin{tabular}{ |l|c|c|c| }
\hline
Dataset &Epinions&Slashdot&Wikipedia\\ \hline
d=10 (exact)&96.36\% &98.00\% &87.69\% \\
d=10 (cplex)&96.17\% & 97.31\% & 86.98\%\\
d=25 (cplex)& 96.20\% &97.52\% &85.60\% \\
METSLib & 93.03\% & 96.79\% & 86.22\%\\\hline
\end{tabular}
\end{table}

In \cite{GuhaKRT04} and \cite{LeskovecHK10} the authors use certain techniques to
test their approaches on unbiased datasets. They use, however, different ways to balance
the dataset and/or results. For instance, \cite{GuhaKRT04} do not change the dataset (Epinions),
but, since the dataset is biased toward positive links, they find the error ratio separately
for positive and negative links, and then average the results. More precisely, they test
the method on a set of randomly sampled edges that naturally contains more 
positive edges. Then they record the error rate on all negative edges, sample randomly
the same number of positive edges (from the test set), find the error rate on them,
and report the mean of the two numbers.

The approach of \cite{LeskovecHK10} is different. Instead of balancing the results they
balance the dataset itself. In order to do that they keep all the negative edges in the
datasets, and then sample the same number of positive edges removing the rest of them.
All the training and testing is done on the modified datasets. Although we have reservations 
about both approaches, we tested our model in these two settings as well. The results 
are shown in Table~\ref{tb5}.
\begin{table} 
\caption{Prediction Accuracy of Balanced Approach}
\label{tb5}
\begin{tabular}{ |l|c|c|c| }
\hline
Dataset & Epinions & Slashdot & Wikipedia\\ \hline
Standard-pd & 85.14 \% & 82.82\% & 62.58\% \\
(averaging the results) &&& \\
of them false negative & 0.84\% & 0.50\% & 2.10\% \\
of them false positive  & 28.89\% & 33.8\% & 72.6\% \\
Standard-pq & 89.36\% & 86.37\% & 76.81\%\\
(balancing the dataset) &&& \\
of them false negative & 6.20\% & 14.37\% & 22.5\% \\
of them false positive & 22.93\% & 15.35\% & 23.90\% \\
All123 (\cite{LeskovecHK10}) & 93.42\% & 93.51\% & 80.21\% \\
EIG (\cite{GuhaKRT04}) & 85.30\% & N/A & N/A \\
\hline
\end{tabular}
\end{table}
\begin{figure*}[ht]
\centering
\epsfig{totalheight=7.5cm, file=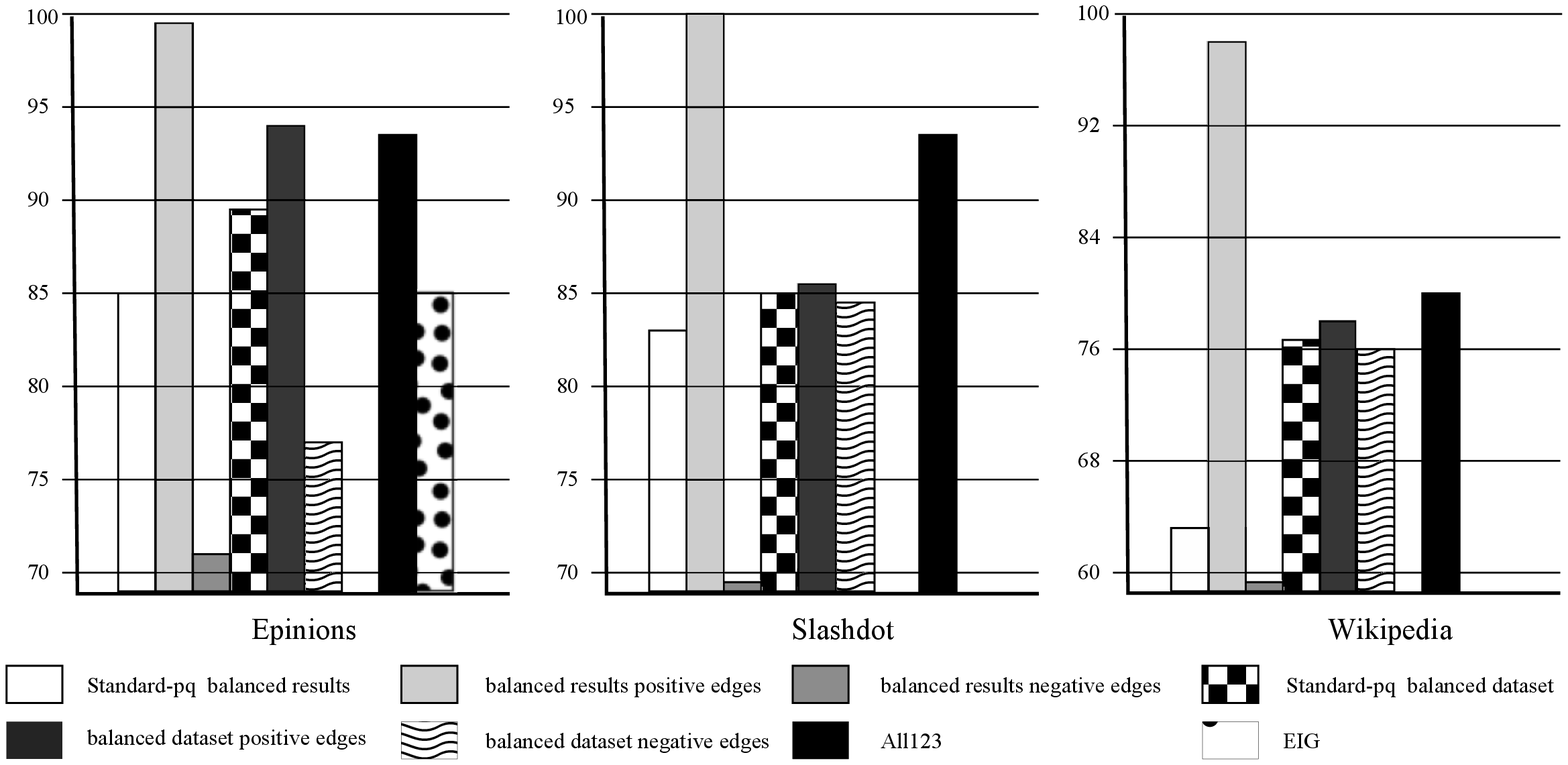}
\caption{Prediction Accuracy of Balanced Approach. Negative Edges 
Bar for Slashdot and Wikipedia are out of Chart.}
\label{fig:balanced}
\end{figure*}

Observe that since our approach is to train the predictor for a particular dataset rather than
finding and tuning up general features as it is done in \cite{GuhaKRT04} and \cite{LeskovecHK10},
and the test datasets are biased toward positive edges, it is natural to expect that predictions
are biased toward positive edges as well. This is clearly seen from Table~\ref{tb5}
(see also Fig.~\ref{fig:balanced}). We therefore 
think that average error rate does not properly reflect the performance of our algorithm. 

In the case of balanced datasets our predictor does not produce biased results, again as expected. 
This, however, is the only case when its performance is worse than some of the previous results.
One way to explain this is to note that density of the dataset is crucial for accurate predictions
made by the quadratic correlation approach. Therefore we had to lower the embeddedness 
threshold used in this part of the experiment to $p=5$, $q=5$, while \cite{LeskovecHK10}
still tests only edges of embeddedness at least 25.

\subsection{Recommender systems dataset}
To test the versatility of the model, we also test it on a completely 
different dataset.
MovieLens \cite{movielens} is a dataset used primarily in the study of 
recommender systems. It contains rating of movies given by users 
who rented movies from a shop or online. Every user gives a rating to
some of the movies by assigning a score from 1 to 5, where higher score
corresponds to higher evaluation of the movie. It therefore can viewed 
as a bipartite graph with users in one part of the bipartition, and 
movies in the other. The version of the dataset we used, MovieLens-100k,
contains approximately 100,000 ratings from 1000 users on 1700
movies. There are also certain density restriction: Every user included in 
the dataset must rate at least 20 movies. 

It is natural to treat users ratings as attitudes of users towards movies. 
Our model, however, cannot work directly with the MovieLens dataset, 
because it requires binary attitudes rather ratings between 1 and 5. Thus,
we convert user ratings into positive and negative attitudes, by introducing 
a negative link every time user's rating is 3 or less, and by introducing a 
positive link if user's rating is 4 or 5. Under such interpretation of scores
the dataset is almost balanced, 44.625\% of its edges are negative.
Predictions are, of course, also made in terms of positive and negative 
links.
 
With the standard values of the parameters: using \emph{Standard-pq} 
option with Cplex, and with $p=15$, $q=20$, $d=10$, the model makes 
about 75\% correct predictions providing about 20\% increase over
the random guess. Although there is a very substantial amount of research 
on recommender systems using MovieLens as a test dataset (see e.g.\ \cite{Sarwar00:analysis,ODonovan05:trust}, it is not 
possible to compare our result against the existing ones, because the 
evaluation measures normally used for recommender systems are quite different; they measure either the success rate in recommending a group of products (movies) or given in terms of estimating user's rating rather than attitude. Nevertheless we can conclude that the method gives a similar advantage over the random choice, as for other datasets. One interesting feature of the (internal) work of
our method is that it finds influences and sets of trusted peers between 
users, although there is no explicit information about such connections.

\section{Conclusion}
We have investigated the link sign prediction problem in online social 
networks with a mixture of both positive and negative relationships. 
We have shown that a better prediction accuracy can be achieved using 
personalized features such as peer opinions. Moreover, the proposed 
model accommodates the dynamic nature of online social networks by 
building a predictor for each individual nodes independently. It enables 
fast updates as the underlaying network evolves over time. 

In the future, we consider possible improvements of the model in two 
directions. First of all, we need to find a better formulation for peer opinions. 
The current formulation is very simple that it either gives an estimation or 
it doesn't estimate. Ideally, we want a formulation that gives an estimation 
along with the confidence level of its estimation. The current choice of the binary 
representation of the problem was determined by 
several factors. Firstly, many more existing algorithms, heuristics, and 
off-the-shelf solvers are available for binary problems. This includes
many readily available Max-SAT solvers. Some solvers,
for example, Cplex, while can be used for non-binary problems, 
produce good results only if the problem satisfies certain conditions. We
experimented with weights $w_{v,x}$ that can take more than just 2 values,
but this often leads to instances that are not positive semidefinite, and Cplex
does not produce any meaningful results. In spite of this we experimented
by allowing various variables of the problem to take more values. 
However, it did not lead to any noticeable improvements of the results.

Secondly, we want to 
build a more sophisticated model that incorporates more information. 
The basic assumption of our model is that users' actions can be determined 
by the opinions of their peers in the network. Yet, as an independent 
individual, we also have our own knowledge and belief. There are also 
external factors that affects our decision making process, such as mood, 
weather, location, and so on. All these information can be used as features 
for our model in the future.

\providecommand{\bysame}{\leavevmode\hbox to3em{\hrulefill}\thinspace}
\providecommand{\MR}{\relax\ifhmode\unskip\space\fi MR }
\providecommand{\MRhref}[2]{%
  \href{http://www.ams.org/mathscinet-getitem?mr=#1}{#2}
}
\providecommand{\href}[2]{#2}

\end{document}